\documentclass[prd,amsmath,amssymb,amsthm,nofootinbib]{revtex4}

\allowdisplaybreaks
\usepackage[normalem]{ulem}
\usepackage{xcolor}
\usepackage[mathscr]{euscript}
\usepackage{bm}
\usepackage{graphicx}
\usepackage{subfigure}
\usepackage[colorlinks=true,linkcolor=red]{hyperref}

\newcommand{\bea}{\begin{eqnarray}}
\newcommand{\eea}{\end{eqnarray}}
\newcommand{\beq}{\begin{equation}}
\newcommand{\eeq}{\end{equation}}
\newcommand{\nn}{\nonumber}
\def\/{\over}

\begin{document}

\title{Entanglement generation for uniformly accelerated atoms assisted by environment-\\
induced interatomic interaction and the loss of the anti-Unruh effect}
\author{Ying Chen, Jiawei Hu\footnote{Corresponding author: jwhu@hunnu.edu.cn},
and Hongwei Yu\footnote{Corresponding author: hwyu@hunnu.edu.cn}}
\affiliation{
Department of Physics and Synergetic Innovation Center for Quantum Effects and Applications, Hunan Normal University, Changsha, Hunan 410081, China}

\begin{abstract}
We study the influence of  the environment-induced interatomic interaction, which is usually neglected, on the entanglement dynamics of two uniformly accelerated atoms coupled with fluctuating massless scalar fields in the Minkowski vacuum.
When the two-atom system is initially prepared in a separable state such that the two atoms are in the ground and excited states respectively, the environment-induced interatomic interaction assists entanglement generation,
in the sense that the parameter space of acceleration and interatomic separation that allows entanglement generation is enlarged, the rate of entanglement generation at the initial time is enhanced, and the maximum of concurrence generated during evolution is increased compared with  when the environment-induced interaction is neglected.
Remarkably, the rate of entanglement generation at the initial time and the maximal concurrence generated during evolution decrease monotonically with the acceleration, in contrast to that they exhibit a nonmonotonic behavior as the acceleration varies when the environment-induced interatomic interaction is neglected. In other words, the anti-Unruh phenomenon in terms of the entanglement generation is deprived of by the consideration of the environment-induced interatomic interaction.

\end{abstract}

\maketitle

\section{Introduction}

Quantum entanglement, which is regarded as one of the most striking features in quantum physics, is at the heart of quantum based novel technologies. However, due to the inevitable  coupling between a quantum system and the fluctuating quantum fields in vacuum, the entanglement of a quantum system can be significantly affected by the vacuum fluctuations it coupled with. This kind of system-environment coupling may cause decoherence and dissipation which in general lead to disentanglement on the one hand, while it may also provide indirect interactions between otherwise separable quantum subsystems which may lead to entanglement generation on the other hand. For inertial quantum systems, phenomena such as the environment-induced entanglement sudden death \cite{Yu1,Yu2}, entanglement generation  \cite{Braun,Kim,ss,Basharov,Jakobczyk,Benatti1,Benatti2,br1,Ficek4,Ficek5,Ficek3,Tana},  and entanglement revival \cite{zf} have been extensively studied.

The situation is more involved when accelerated quantum systems are concerned, since the vacuum state of a quantum field is observer dependent. A uniformly accelerated observer perceives the Minkowski vacuum as a thermal bath of Rindler particles at a temperature proportional to the proper acceleration, which is known as the Unruh effect \cite{Unruh,sa,pc,lc}. Therefore, it is of interest to study the entanglement dynamics of a uniformly accelerated two-atom system. At this point, let us note that the asymptotic entanglement of a two-atom system with a vanishing separation interacting with fluctuating scalar fields in the Minkowski vacuum has been studied in Ref. \cite{Benatti3}, and this work has been generalized to the case with the presence of a reflecting plate \cite{Zhang}, and in a background with a nonzero temperature \cite{Lima20}.
Apart from the asymptotic entanglement, the time evolution of entanglement for accelerated atoms has also been examined \cite{Landulfo,Hu,Yang,Cheng}. In particular, it has been found that the maximal concurrence generated during evolution for an initially separable two-atom system may increase with acceleration, in contrast to the expectation that it should decrease monotonically with the Unruh temperature \cite{Hu,zhou21}. This phenomenon is dubbed as the anti-Unruh phenomenon in terms of the entanglement generated in Ref. \cite{zhou21}, in analogy to the anti-Unruh phenomenon in terms of the excitation rate \cite{anti1,anti2}, i.e., the excitation rate of a uniformly accelerated detector may decrease with the Unruh temperature in certain cases.

Actually, from the Gorini-Kossakowski-Lindblad-Sudarshan master equation  \cite{Kossakowski,Lindblad,Breure} which describes the evolution of an open quantum system, it is clear that the role the environment played is twofold: One is  the decoherence and dissipation caused by the environment which  is described by a nonunitary term, and the other is the  unitary evolution described by a unitary term  which contains the environment-induced energy shift. For a two-atom system, the  energy shift term includes the Lamb shifts of the individual atoms, and an environment-induced interatomic interaction.
In most of the existing literature dealing with the entanglement dynamics for accelerated atoms \cite{Benatti3,Lima20,Zhang,Hu,Yang,Cheng,Landulfo},  the environment-induced interatomic interaction is neglected. However, as we will show in detail in the present paper, the environment-induced interatomic interaction may have a significant effect on the entanglement dynamics, even when the interatomic separation is larger than the  atomic transition wavelength such that the environment-induced interatomic interaction is small.  Moreover, the entanglement generation can be assisted by the environment-induced interaction in several aspects, and the anti-Unruh phenomenon in terms of the entanglement generated disappears when the environment-induced  interatomic interaction is taken into account. In the present paper, natural units with $\hbar=c=k_B=1$ are used, where $c$ is the speed of light, $\hbar$  the reduced Planck constant, and $k_B$  the Boltzmann constant.

\section{The Master Equation}

We study the entanglement dynamics of an open quantum system composed of two uniformly accelerated two-level atoms, which are weakly coupled to fluctuating massless scalar fields in vacuum.
The Hamiltonian of the two atoms $H_{S}$ can be expressed as
\begin{equation}\label{hs}
H_{S}=\frac{\omega}{2}\sigma_{3}^{(1)}+\frac{\omega}{2}\sigma_{3}^{(2)}.
\end{equation}
Here $\sigma^{(1)}_{i}=\sigma_{i}\otimes\sigma_{0}$, $\sigma^{(2)}_{i}=\sigma_{0}
\otimes\sigma_{i} $ are operators of atom 1 and 2 respectively, with $\sigma_{i}~(i=1,2,3)$ being the Pauli matrices, and $\sigma_{0}$ the $2\times2$ unit matrix. We assume that the transition frequencies of the two atoms are the same, and label it as $\omega$.
The interaction Hamiltonian $H_{I}$ between the atoms and the vacuum scalar fields can be written in analogy to the atom-light interaction as \cite{Audretsch1994}
\beq
H_{I}=\mu[\sigma^{(1)}_{2}\Phi(t,x_{1})+\sigma^{(2)}_{2}\Phi(t,x_{2})].
\eeq
Here $\mu$ is the coupling constant which is assumed to be small.

In the Born-Markov approximation, the master equation describing the dissipative dynamics of the two-atom subsystem can be written in the  Gorini-Kossakowski-Lindblad-Sudarshan form as \cite{Kossakowski,Lindblad,Breure}
\beq\label{master1}
\frac{\partial\rho(\tau)}{\partial\tau}=-i[H_{\rm eff},\rho(\tau)]+
\mathcal{D}[\rho(\tau)],
\eeq
where
\beq\label{master2}
H_{\rm eff}=H_{S}-\frac{i}{2}\sum^{2}_{\alpha,\beta=1}\sum^{3}_{i,j=1}
H_{ij}^{(\alpha\beta)}\sigma_{i}^{(\alpha)}\sigma_{j}^{(\beta)},
\eeq
and
\beq\label{master3}
\mathcal{D}[\rho(\tau)]=\frac{1}{2}\sum^{2}_{\alpha,\beta=1}\sum^{3}_{i,j=1}
C_{ij}^{(\alpha\beta)}[2\sigma_{j}^{(\beta)}\rho\sigma_{i}^{(\alpha)}-
\sigma_{i}^{(\alpha)}\rho\sigma_{j}^{(\beta)}-\rho\sigma_{i}^{(\alpha)}
\sigma_{j}^{(\beta)}].
\eeq
From the master equation \eqref{master1},  it is clear that the environment leads to decoherence and dissipation  described by the dissipator $\mathcal{D}[\rho(\tau)]$  such that the evolution of the quantum system is nonunitary on the one hand, and it also gives rise to a modification of the  unitary evolution term which incarnates  in  the Hamiltonian $H_{\rm eff}$   on the other hand.
The coefficients $C_{ij}^{(\alpha\beta)}$ in the dissipator \eqref{master3} can be written as
\beq
C_{ij}^{(\alpha\beta)}=A^{(\alpha\beta)}\delta_{ij}-iB^{(\alpha\beta)}
\epsilon_{ijk}\delta_{3k}-A^{(\alpha\beta)}\delta_{3i}\delta_{3j},
\eeq
where
\bea\label{pf2}
\nn A^{(\alpha\beta)}=\frac{\mu^{2}}{4}[\mathcal{G}^{(\alpha\beta)}(\omega)
+\mathcal{G}^{(\alpha\beta)}(-\omega)],\\
B^{(\alpha\beta)}=\frac{\mu^{2}}{4}[\mathcal{G}^{(\alpha\beta)}(\omega)
-\mathcal{G}^{(\alpha\beta)}(-\omega)].
\eea
In the above expressions
\beq
\mathcal{G}^{(\alpha\beta)}(\lambda)=\int^{\infty}_{-\infty}d\Delta\tau
e^{i\lambda\Delta\tau}\langle\Phi(\tau,x_{\alpha})\Phi(\tau',x_{\beta})\rangle,
\eeq
is the Fourier transform of the scalar field correlation function $\langle\Phi(\tau,x_{\alpha})\Phi(\tau',x_{\beta})\rangle$.
Similarly, $H_{ij}^{(\alpha\beta)}$ can be obtained by replacing the Fourier transform $\mathcal{G}^{(\alpha\beta)}$ with the Hilbert transform $\mathcal{K}^{(\alpha\beta)}$, i.e.,
\beq
\mathcal{K}^{(\alpha\beta)}(\lambda)=\frac{P}{\pi i}\int^{\infty}_{-\infty}d\omega
\frac{\mathcal{G}^{(\alpha\beta)}(\omega)}{\omega-\lambda},
\eeq
with  $P$ being the principal value.
Then, it is found that the effective Hamiltonian ${ H_{\rm eff}}$ can be divided into two parts, i.e., ${ H_{\rm eff}}=\tilde{H_{s}}+{ H^{(12)}_{\rm eff}}$.  The first part $\tilde{H_{s}}$ describes the  renormalization of the transition frequencies, i.e., the Lamb shift of each individual atom. It takes the same form as  Eq. (\ref{hs}), but with a redefined energy level spacing
\bea
\tilde{\omega}=\omega-\frac{i\mu^{2}}{2}[\mathcal{K}^{(11)}(\omega)-
\mathcal{K}^{(11)}(-\omega)].
\eea
In the following, this term will be neglected since it can be regarded as a rescaling of the gap of the  energy level.
The second part ${ H^{(12)}_{\rm eff}}$ describes the environment-induced coupling between the two atoms, which takes the form
\bea
H^{(12)}_{\rm eff}=-\sum_{i,j=1}^{3}\Omega_{ij}^{(12)}(\sigma_{i}\otimes\sigma_{j}),
\eea
where
\bea
\Omega_{ij}^{(12)}=\frac{i\mu^{2}}{4}\{[\mathcal{K}^{(12)}(\omega)+\mathcal{K}^{(12)}(-\omega)]\delta_{ij}-
[\mathcal{K}^{(12)}(\omega)+\mathcal{K}^{(12)}(-\omega)]\delta_{3i}\delta_{3j}\}.
\eea
Therefore, the master equation can be rewritten in the following form,
\bea
\nn
\frac{\partial\rho(\tau)}{\partial\tau}&=&-i\tilde{\omega}\sum_{\alpha=1}^{2}\,
[\sigma_{3}^{(\alpha)},\rho(\tau)]+i
\sum_{i,j=1}^{3}\Omega_{ij}^{(12)}[\sigma_{i}\otimes\sigma_{j},\rho(\tau)]\\
&&+\frac{1}{2}\sum^{2}_{\alpha,\beta=1}\sum^{3}_{i,j=1}
C_{ij}^{(\alpha\beta)}[2\sigma_{j}^{(\beta)}\rho\sigma_{i}^{(\alpha)}-
\sigma_{i}^{(\alpha)}\rho\sigma_{j}^{(\beta)}-\rho\sigma_{i}^{(\alpha)}
\sigma_{j}^{(\beta)}].
\eea

\section{Entanglement dynamics for accelerated atoms}
In this section, we investigate the entanglement dynamics of the two-atom system, and focus on how the entanglement dynamics is affected by the environment-induced interatomic interaction.
We assume that the two atoms are accelerated uniformly with the same acceleration along the $x$ axis, and are separated with a distance $L$ along the $z$ axis.  So, the trajectories can be written in the following form,
\bea\label{traj15}
\nn t_{1}(\tau)=\frac{1}{a}\sinh(a\tau),\;\;x_{1}(\tau)=\frac{1}{a}\cosh(a\tau),
\;\;y_{1}(\tau)=0,\;\;z_{1}(\tau)=0,\\
t_{2}(\tau)=\frac{1}{a}\sinh(a\tau),\;\;x_{2}(\tau)=\frac{1}{a}\cosh(a\tau),
\;\;y_{2}(\tau)=0,\;\;z_{2}(\tau)=L.
\eea
Plugging the trajectories Eq. \eqref{traj15} into the Wightman function of the massless scalar fields in the Minkowski vacuum
\bea
\langle\Phi(t,x_{\alpha})\Phi(t',x_{\beta})\rangle=-\frac{1}{4\pi^{2}}
\frac{1}{(t-t'-i\epsilon)^{2}-(x-x')^{2}-(y-y')^{2}-(z-z')^{2}},
\eea
it can be found that the Fourier transforms of the correlation functions are
\bea
\mathcal{G}^{(11)}(\omega)&=&\mathcal{G}^{(22)}(\omega)=\frac{1}{2\pi}\frac{\omega}
{1-e^{-\frac{2\pi\omega}{a}}},\\
\mathcal{G}^{(12)}(\omega)&=&\mathcal{G}^{(21)}(\omega)=\frac{1}{2\pi}\frac{\omega}
{1-e^{-\frac{2\pi\omega}{a}}}
\frac{\sin(\frac{2\omega}{a}\sinh^{-1}\frac{a L}{2})}
{\omega L\sqrt{1+a^{2}L^{2}/4}}.
\eea
From this we obtain
\bea\label{cij19}
&&C_{ij}^{(11)}=C_{ij}^{(22)}=A_{1}\delta_{ij}-i B_{1}\epsilon_{ijk}\delta_{3k}-A_{1}\delta_{3i}\delta_{3j},\\
&&C_{ij}^{(12)}=C_{ij}^{(21)}=A_{2}\delta_{ij}-i B_{2}\epsilon_{ijk}\delta_{3k}-A_{2}\delta_{3i}\delta_{3j},\\
&&\Omega^{(12)}_{ij}=D\delta_{ij}-D\delta_{3i}\delta_{3j},\label{cij21}
\eea
where
\bea
\nn A_{1}&=&\frac{\Gamma_{0}}{4}\coth\frac{\pi\omega}{a},\\
\nn A_{2}&=&\frac{\Gamma_{0}}{4}
\frac{\sin(\frac{2\omega}{a}\sinh^{-1}\frac{a L}{2})}
{\omega L\sqrt{1+a^{2}L^{2}/4}}
\coth\frac{\pi\omega}{a},\\
\nn B_{1}&=&\frac{\Gamma_{0}}{4},\\
\nn B_{2}&=&\frac{\Gamma_{0}}{4}
\frac{\sin(\frac{2\omega}{a}\sinh^{-1}\frac{a L}{2})}
{\omega L\sqrt{1+a^{2}L^{2}/4}},\\
D&=&\frac{\Gamma_{0}}{4}\frac{\cos(\frac{2\omega}{a}\sinh^{-1}\frac{a L}{2})}
{\omega L\sqrt{1+a^{2}L^{2}/4}},
\eea
and $\Gamma_{0}=\frac{\mu^{2}\omega}{2\pi}$ is the spontaneous emission rate for an inertial atom in the Minkowski vacuum.

For convenience, we work in the coupled basis $\{|G\rangle=|00\rangle,|A\rangle=\frac{1}{\sqrt{2}}(|10\rangle-|01\rangle),
|S\rangle=\frac{1}{\sqrt{2}}(|10\rangle+|01\rangle),|E\rangle=|11\rangle\}$, and obtain the following time evolution equations  of the  density matrix elements \cite{Ficek}
\bea\label{pf2}
\nn \rho_{GG}'&=&-4(A_{1}-B_{1})\rho_{GG}+2(A_{1}+B_{1}-A_{2}-B_{2})\rho_{AA}
+2(A_{1}+B_{1}+A_{2}+B_{2})\rho_{SS},\\
\nn \rho_{EE}'&=&-4(A_{1}+B_{1})\rho_{EE}+2(A_{1}-B_{1}-A_{2}+B_{2})\rho_{AA}
+2(A_{1}-B_{1}+A_{2}-B_{2})\rho_{SS},\\
\nn \rho_{AA}'&=&-4(A_{1}-A_{2})\rho_{AA}+2(A_{1}-B_{1}-A_{2}+B_{2})\rho_{GG}
+2(A_{1}+B_{1}-A_{2}-B_{2})\rho_{EE},\\
\nn \rho_{SS}'&=&-4(A_{1}+A_{2})\rho_{SS}+2(A_{1}-B_{1}+A_{2}-B_{2})\rho_{GG}
+2(A_{1}+B_{1}+A_{2}+B_{2})\rho_{EE},\\
\nn \rho_{AS}'&=&-4(A_{1}+i D)\rho_{AS},\;\;\;\;\;\;\;\;\;\;\;\;\;\;\;\;\;\;\;\;\;
\;\;\;\;\;\;\;\;\;\;\;\;\;\;\rho_{SA}'=-4(A_{1}-i D)\rho_{SA},\\
\rho_{GE}'&=&-4A_{1}\rho_{GE},\;\;\;\;\;\;\;\;\;\;\;\;\;\;\;\;\;\;\;\;\;
\;\;\;\;\;\;\;\;\;\;\;\;\;\;\;\;\;\;\;\;\;\;\;\;\;\rho_{EG}'=-4A_{1}\rho_{EG},
\eea
where $\rho_{IJ}=\langle I|\rho|J\rangle,I,J\in\{G,E,A,S\}$ and $\rho_{IJ}'=\frac{\partial\rho_{IJ}(\tau)}{\partial\tau}$. If we assume that the initial density matrix is of the  X form, i.e., the nonzero elements are arranged along the diagonal and antidiagonal of the density matrix,
then the X structure will be preserved during evolution, and the concurrence \cite{Wootters} which quantifies the amount of quantum entanglement of a two-atom system in the X state is found to be~\cite{Ficek2}
\bea\label{pfc}
C[\rho(\tau)]=\max\{0,K_{1}(\tau),K_{2}(\tau)\},
\eea
where
\bea\label{pfk}
\nn K_{1}(\tau)&=&\sqrt{[\rho_{AA}(\tau)-\rho_{SS}(\tau)]^{2}-[\rho_{AS}(\tau)-\rho_{SA}(\tau)]^{2}}-
2\sqrt{\rho_{GG}(\tau)\rho_{EE}(\tau)},\\
K_{2}(\tau)&=&2|\rho_{GE}(\tau)|-\sqrt{[\rho_{AA}(\tau)+\rho_{SS}(\tau)]^{2}-
[\rho_{AS}(\tau)+\rho_{SA}(\tau)]^{2}}.
\eea

\subsection{Entanglement generation }

In this part, we investigate the entanglement generation between two initially separable uniformly accelerated atoms, and focus on the role played by the environment-induced interatomic interaction.
From Eq. (\ref{pf2}), it is clear that the only density matrix elements that are affected by the environment-induced interatomic interaction in the coupled basis are $\rho_{AS}$ and $\rho_{SA}$. If an initial state is chosen such that $\rho_{AS}(0)=\rho_{SA}(0)=0$, then $\rho_{AS}(\tau)$ and $\rho_{SA}(\tau)$ remain zero, and therefore the environment-induced interatomic interaction has no effect on the entanglement dynamics.
In the following discussions, we assume that the initial state of the two-atom system is $|10\rangle$ [$\rho_{AS}(0)=\rho_{SA}(0)=\frac{1}{2}$].

\subsubsection{The condition for entanglement generation}

\begin{figure}[htbp]
  \centering
  \includegraphics[width=0.35\textwidth]{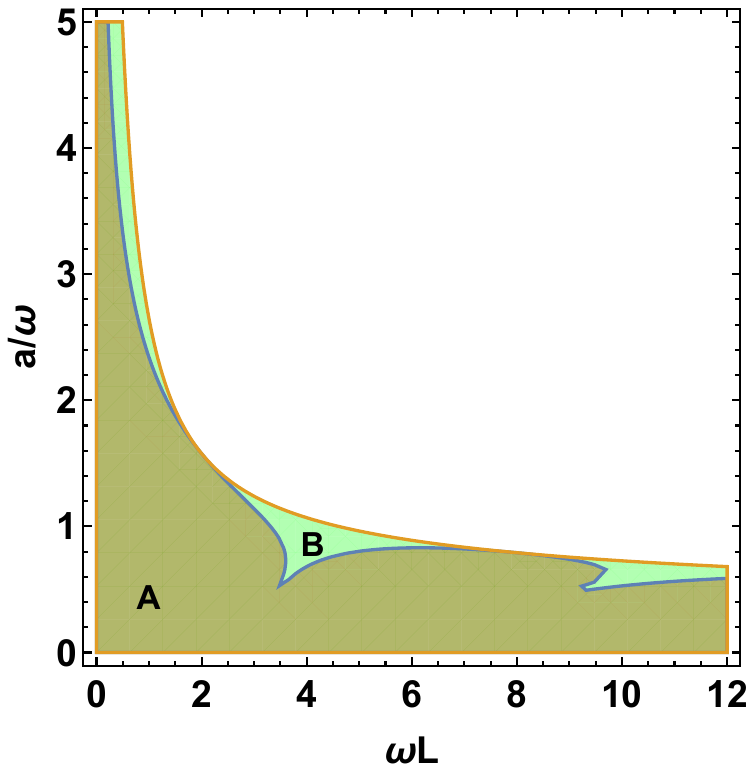}
  \caption{Parameter region $(\omega L, a/\omega)$ within which entanglement generation  is possible for a uniformly accelerated two-atom system initially prepared in $|10\rangle$ with (Regions A and B) and without (Region A) the environment-induced interatomic interaction.   }\label{a1}
\end{figure}

First, we investigate whether entanglement can be generated at the neighborhood of the initial time with the help of the evolution equations (\ref{pf2}) and the expression of concurrence \eqref{pfc}.
When the initial state is chosen as $|10\rangle$, it is obvious from Eq. (\ref{pf2}) that $\rho_{GE}(\tau)\equiv0$, so $K_{2}(\tau)<0$, and the concurrence $C[\rho(\tau)]= \max\{0,K_{1}(\tau)\}$.
Moreover, since $K_{1}(0)=0$, entanglement can be created at the neighborhood of the initial time $\tau=0$ when $K_{1}'(0)
>0$.
Direct calculations show that
\bea\label{k1-28}
{K}_{1}'(0)=4\sqrt{A_{2}^{2}+D^{2}}-4\sqrt{A_{1}^{2}-B_{1}^{2}}.
\eea
Therefore, the condition for entanglement generation at the neighborhood of the initial time is
\bea\label{pf8}
A_{2}^{2}+D^{2}>A_{1}^{2}-B_{1}^{2}.
\eea
Compared with the result when the environment-induced interaction is neglected ($D=0$), entanglement is more likely to be generated when the environment-induced interaction is taken into account.
In Fig. \ref{a1}, we show the parameter region $(\omega L, a/\omega)$ within which entanglement can be generated for a pair of uniformly accelerated atoms. It is shown that the parameter space for entanglement generation is enlarged when the environment-induced interaction between the atoms is taken into account.

\subsubsection{The rate of entanglement generation}

Now, we further investigate how the rate of entanglement generation at the initial time $C'(0)$ is dependent on the acceleration  and the interatomic separation. Before explicit numerical calculations, we note that it is obvious from Eq. \eqref{k1-28} that $C'(0)$ is always larger when the environment-induced interaction is considered ($D\neq0$) compared with that when the environment-induced interaction is neglected ($D=0$). In this sense, the environment-induced interaction assists entanglement generation.

\begin{figure}[htbp]
  \centering
  \includegraphics[width=0.32\textwidth]{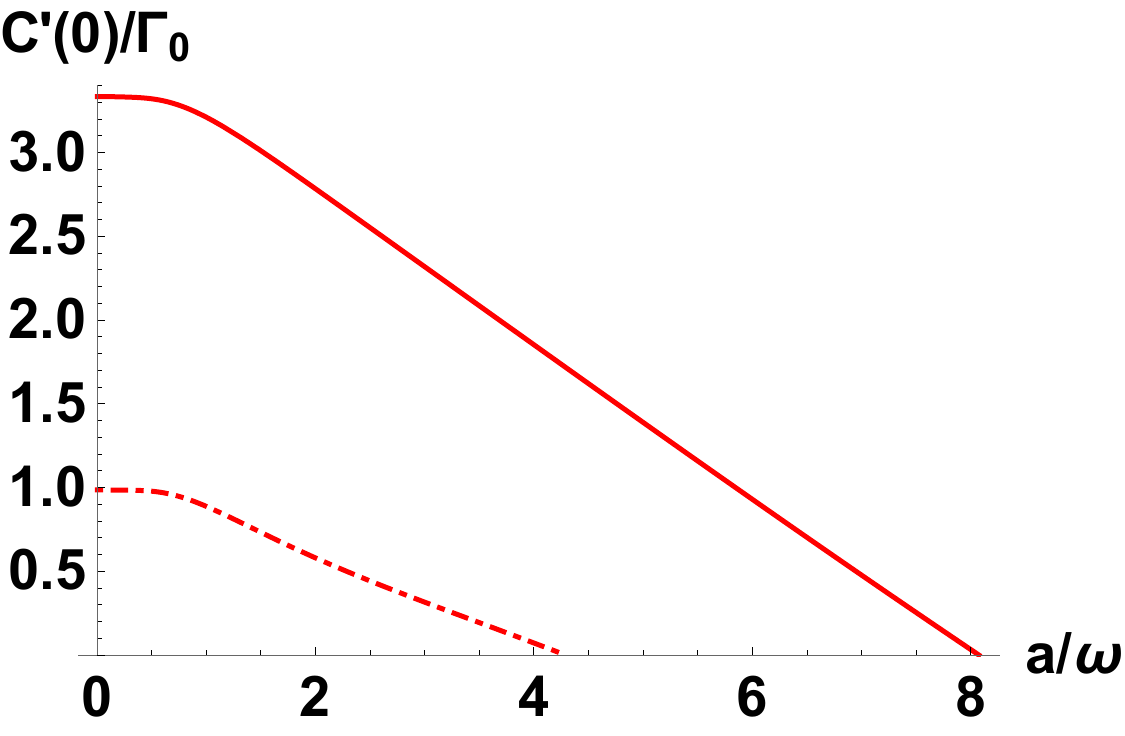}
  \includegraphics[width=0.32\textwidth]{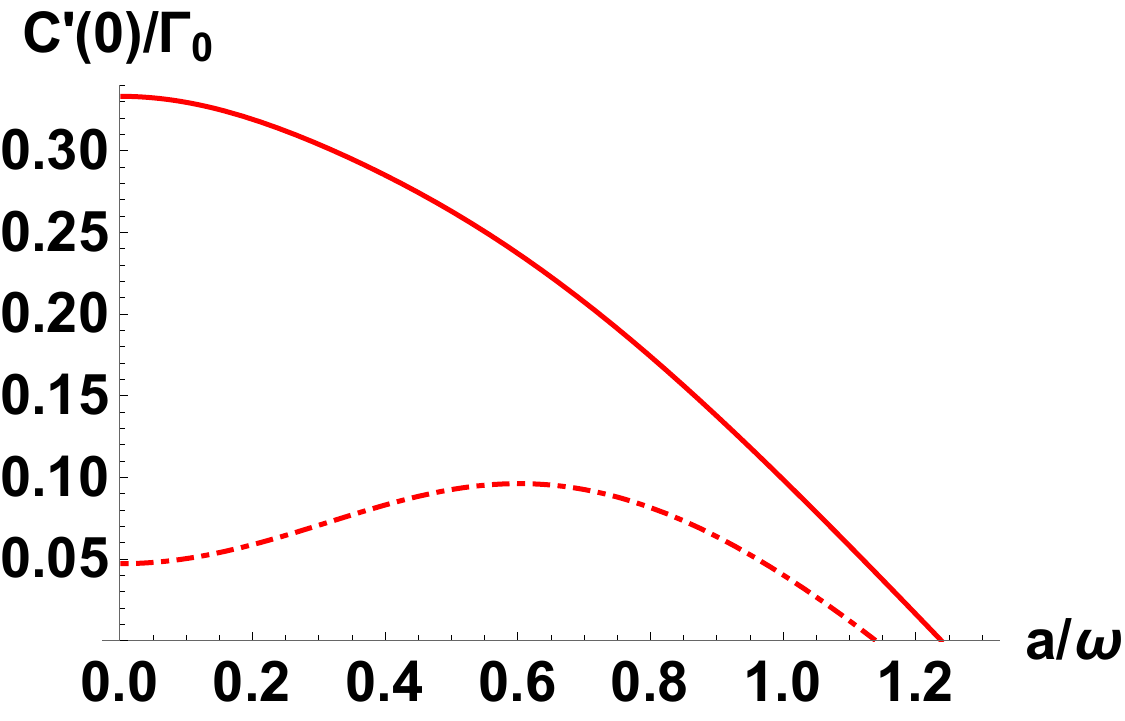}
  \includegraphics[width=0.32\textwidth]{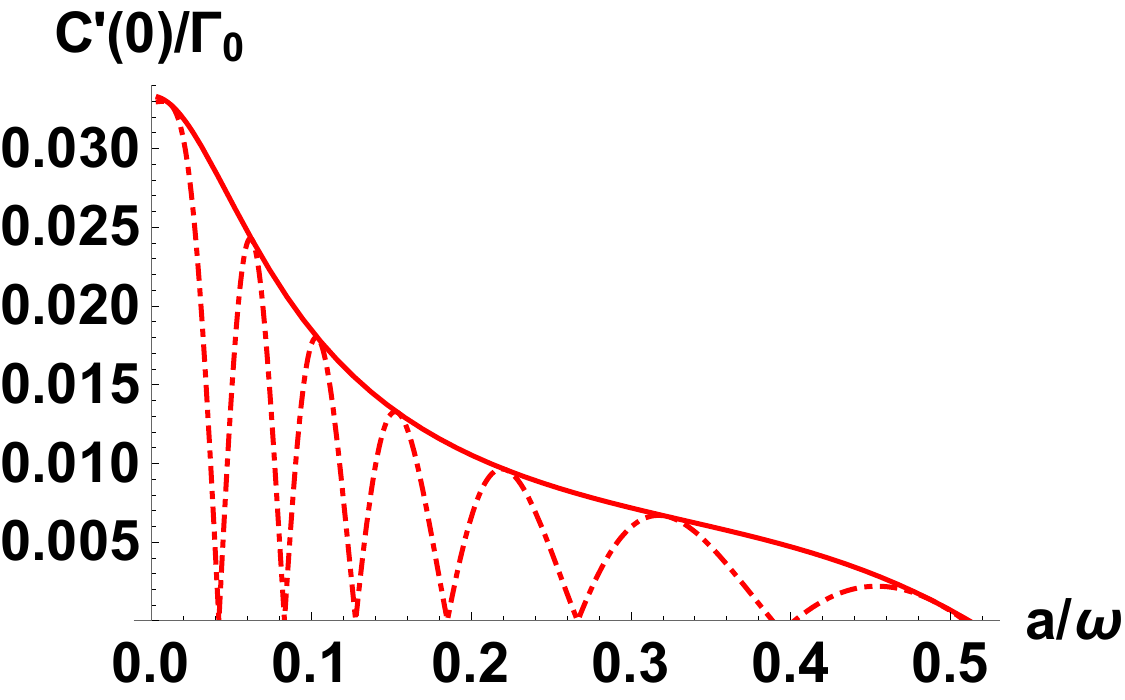}
  \caption{$C'(0)$ as a function of $a/\omega$ with $\omega L=3/10$ (left), $\omega L=3$ (middle), and $\omega L=30$ (right), for uniformly accelerated atoms with (solid) and without (dot-dashed)  the environment-induced interatomic interaction.   }\label{ka}
\end{figure}

In Fig. \ref{ka}, we plot the rate of entanglement generation at the initial time $C'(0)$ as a function of acceleration for a given interatomic separation. It is shown that, $C'(0)$ decreases monotonically with acceleration for any given separation when the  environment-induced interatomic interaction is considered, while for two-atom systems with a separation  comparable to or much larger than the transition wavelength, the relation between $C'(0)$ and acceleration is nonmonotonic  when the environment-induced interatomic interaction is neglected.
That is, even when the interatomic separation is large compared with the atomic transition wavelength such that the environment-induced interatomic interaction is small, the consideration of the  interatomic interaction may still cause significant qualitative changes.

\begin{figure}[htbp]
  \centering
  \includegraphics[width=0.32\textwidth]{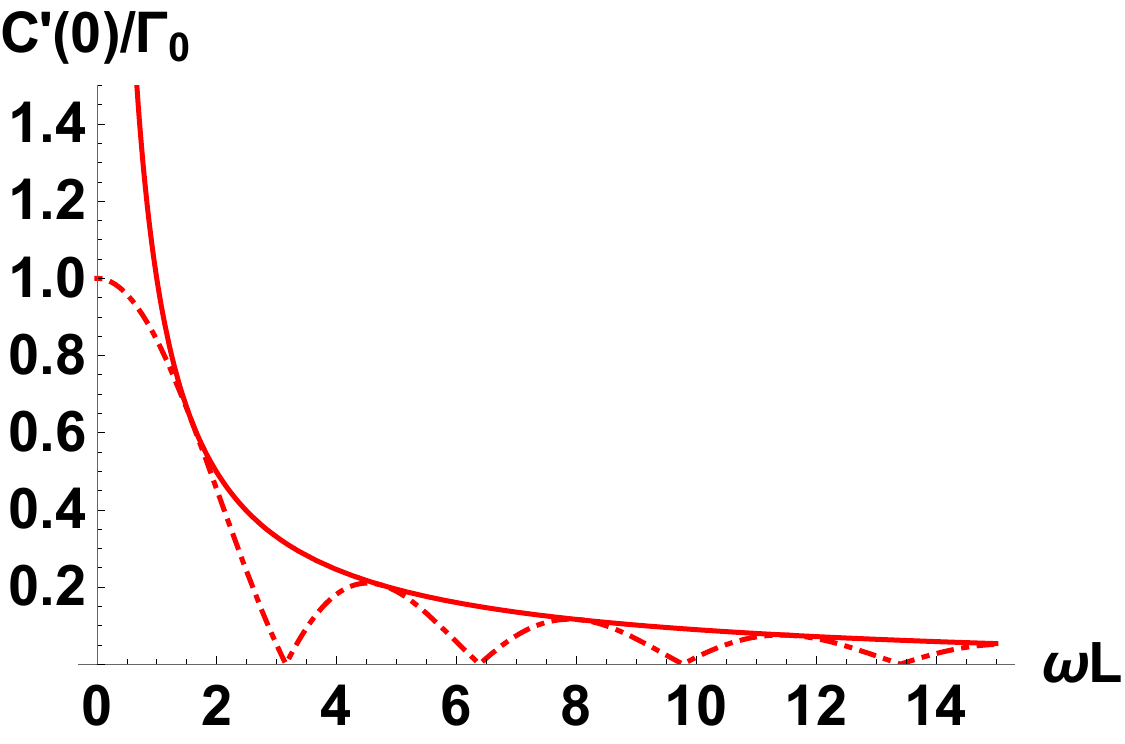}
  \includegraphics[width=0.32\textwidth]{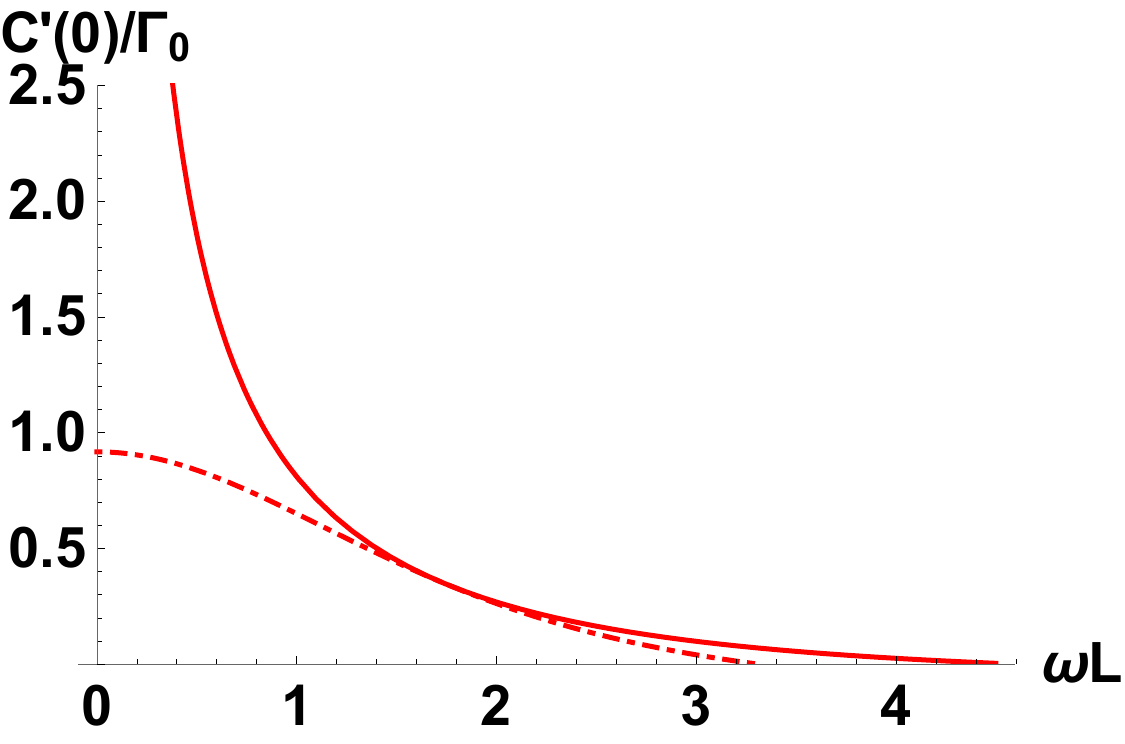}
  \includegraphics[width=0.32\textwidth]{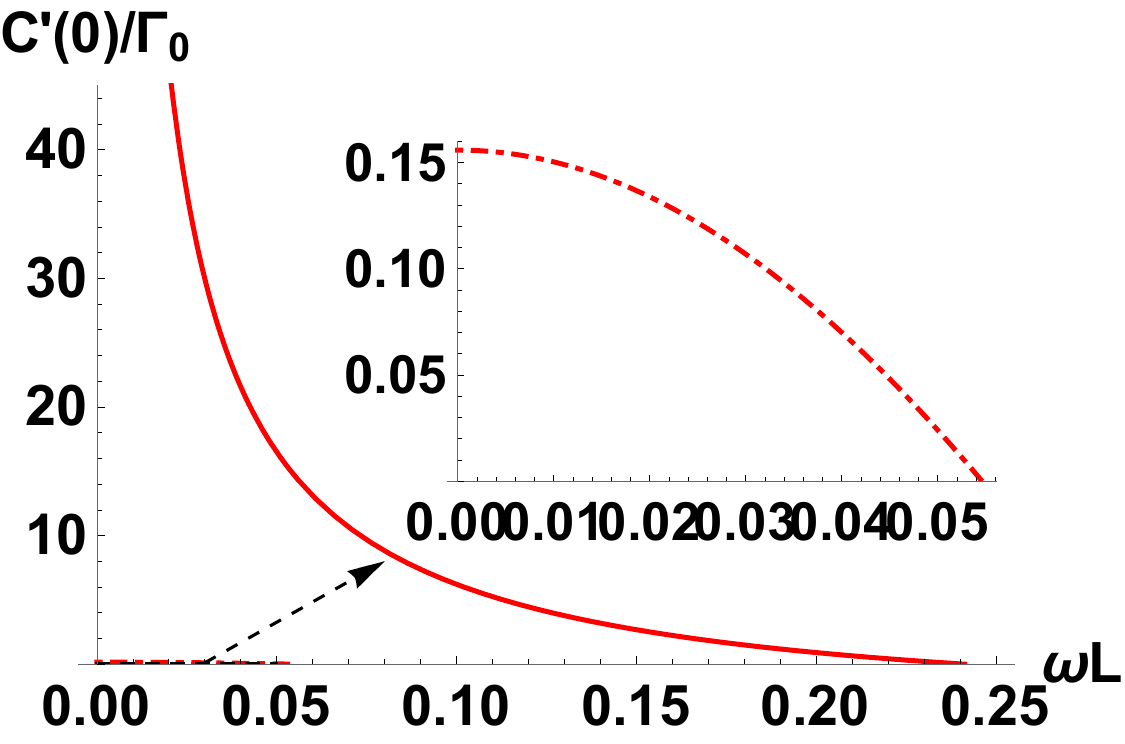}
  \caption{$C'(0)$ as a function of $\omega L$ with $a/\omega=1/10$ (left) , $a/\omega=1$ (middle), and $a/\omega=10$ (right), for uniformly accelerated atoms with (solid) and without (dot-dashed)  the environment-induced interatomic interaction.  }\label{kl}
\end{figure}

Next, we study the rate of entanglement generation at the initial time $C'(0)$ as a function of the interatomic separation.
First of all,  when the interatomic separation is small compared with the transition wavelength, $C'(0)$ is greatly enhanced when the environment-induced interaction is considered, due to a large environment-induced interaction between atoms with a small separation.
In Fig. \ref{kl}, it is shown that $C'(0)$ decreases monotonically with the interatomic separation when the environment-induced interatomic interaction is considered.
In contrast, when the environment-induced interatomic interaction is ignored, $C'(0)$ shows an oscillatory behavior with interatomic separation, when the thermal frequency corresponding to the Unruh temperature is much smaller compared with the atomic transition frequency.
Remarkably,  when the thermal frequency  corresponding to the Unruh temperature is much larger compared with the transition frequency of the atoms,  $C'(0)$ can be significantly enhanced when the environment-induced interatomic interaction is considered.

\subsubsection{The evolution process of entanglement}

\begin{figure}[htbp]
  \centering
  \includegraphics[width=0.4\textwidth]{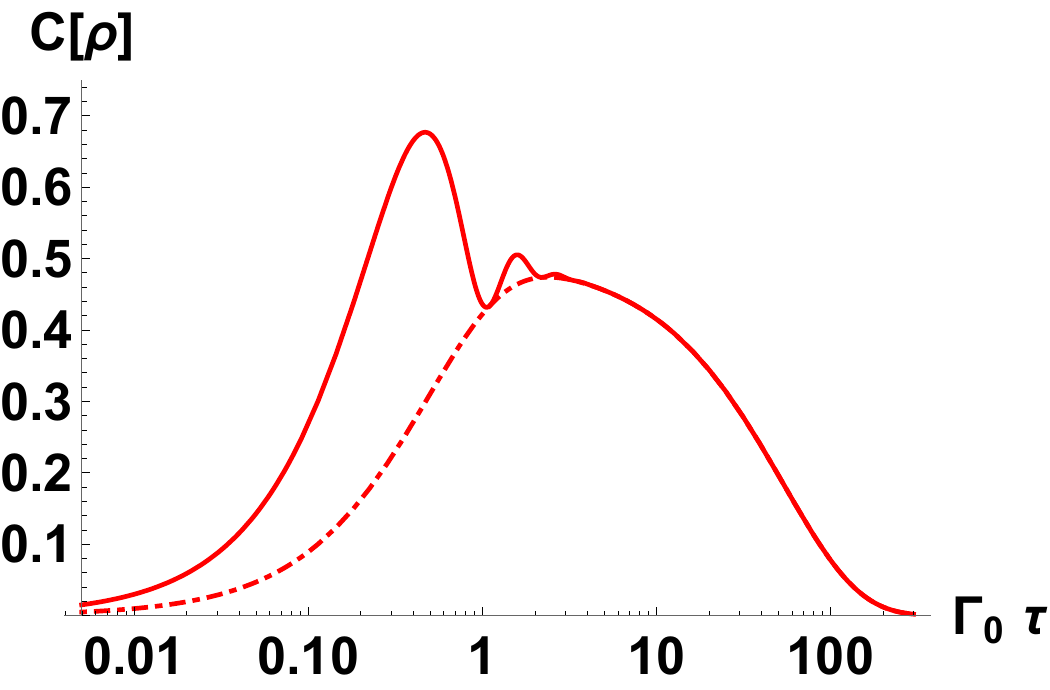}
  \caption{Time evolution of concurrence for uniformly accelerated atoms  with (solid) and without (dot-dashed)  the environment-induced  interatomic interaction initially prepared in $|10\rangle$. Here
$a/\omega=1/10$, and $\omega L = 1/2$.}\label{pg1}
\end{figure}

Now we study the evolution process of entanglement for a two-atom system initially prepared in the state $|10\rangle$. Form Eq. (\ref{pf2}), the time evolution of the density matrix elements $\rho_{AS}(\tau)$
can be solved as
\bea
\rho_{AS}(\tau)=\rho_{SA}^*(\tau)=\frac{1}{2} e^{-4(A_{1}+iD)\tau}.
\eea
Then the concurrence can be calculated with the help of Eqs. (\ref{pfc})-(\ref{pfk}) as $C[\rho(\tau)]=\max\{0,K_{1}(\tau)\}$, where
\bea\label{kk}
K_{1}(\tau)=\sqrt{[\rho_{AA}(\tau)-\rho_{SS}(\tau)]^{2}+\sin^{2}(4D\tau)e^{-4A_{1}\tau}}-2\sqrt{\rho_{GG}(\tau)\rho_{EE}(\tau)}.
\eea
According to Eq. (\ref{kk}), there exists an extra term $\sin^{2}(4D\tau)\,e^{-4A_{1}\tau}$ when the  environment-induced  interatomic interaction is taken into account. As a result, there is an oscillatory term  in the time evolution of concurrence, and the oscillation is damped during evolution, so the asymptotic concurrence will not be affected by the environment-induced interatomic interaction. A typical evolution process of concurrence is shown in Fig. \ref{pg1}. It is shown that concurrence increases from 0 to its maximum, then oscillates with a damping amplitude and vanishes in the asymptotic state.

\subsubsection{The maximum of concurrence generated during evolution}
\begin{figure}[htbp]
  \centering
   \includegraphics[width=0.32\textwidth]{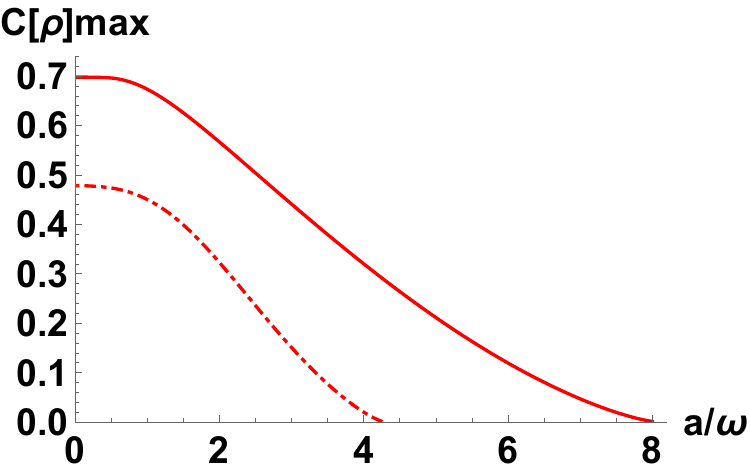}
  \includegraphics[width=0.32\textwidth]{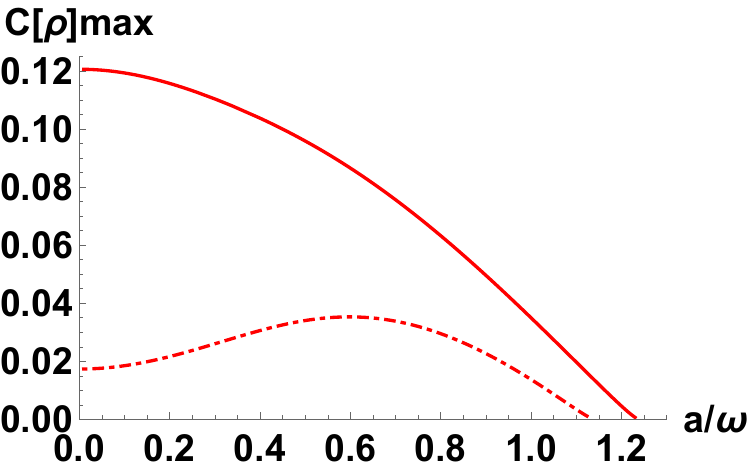}
  \includegraphics[width=0.32\textwidth]{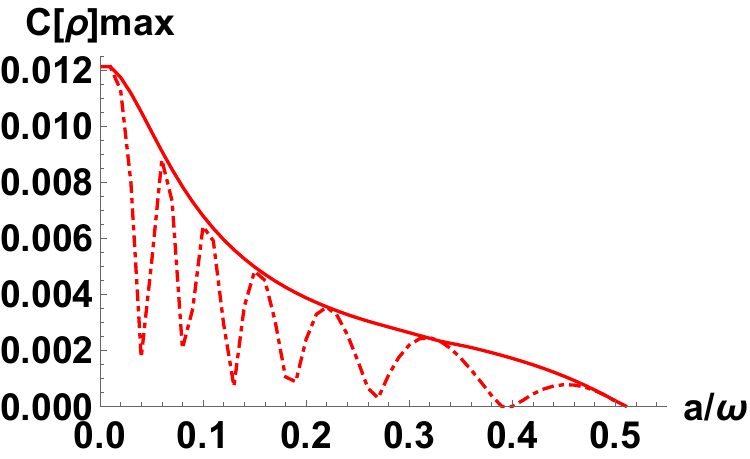}
  \caption{The maximum of concurrence during evolution for uniformly accelerated atoms with (solid) and without (dot-dashed)  the environment-induced interatomic interaction initially prepared in $|10\rangle$, with $\omega L=3/10$ (left), $\omega L=3$ (middle), and $\omega L=30$ (right).}\label{c1}
\end{figure}

\begin{figure}[htbp]
  \centering
  \includegraphics[width=0.32\textwidth]{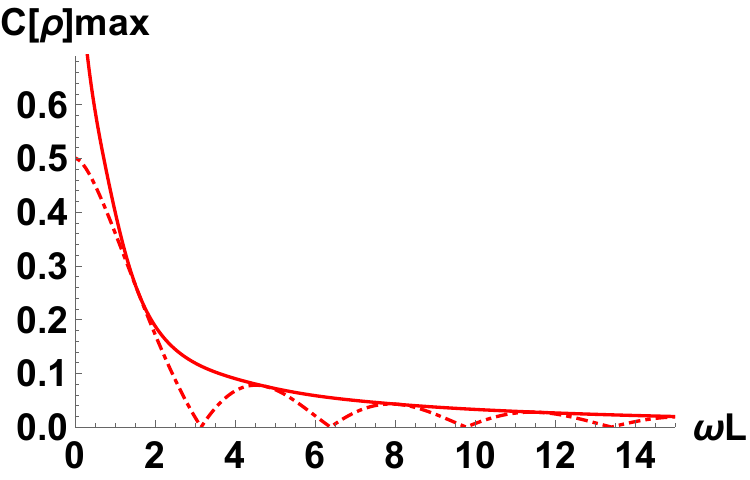}
  \includegraphics[width=0.32\textwidth]{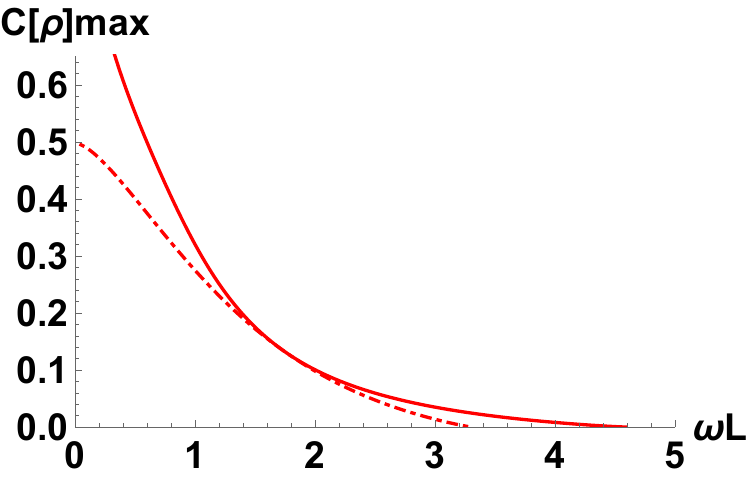}
  \includegraphics[width=0.32\textwidth]{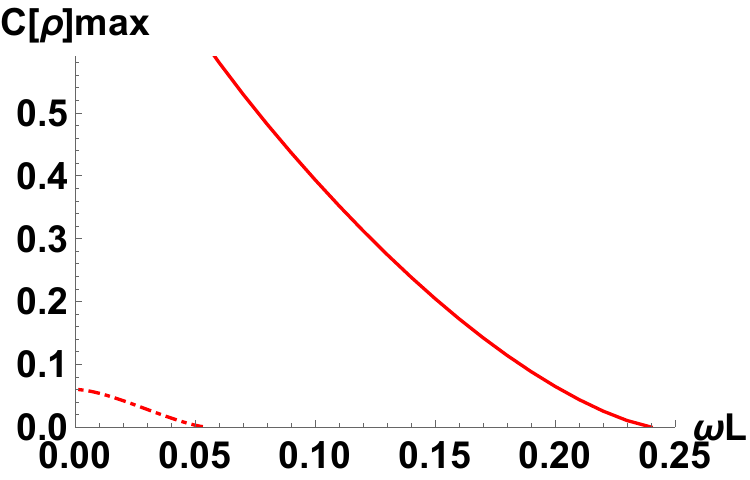}
  \caption{The maximum of concurrence during evolution for uniformly accelerated atoms with (solid) and without (dot-dashed)  the environment-induced interatomic interaction initially prepared in $|10\rangle$, with $a/\omega=1/10$ (left), $a/\omega=1$ (middle), and $a/\omega=10$ (right).}\label{c2}
\end{figure}

In the following, we address the issue of the maximum of concurrence generated during evolution. As shown in Figs. \ref{c1} and \ref{c2}, the maximum of concurrence generated during evolution is always larger when the environment-induced interatomic interaction is taken into account, compared with that in the case when the environment-induced interatomic interaction is neglected.
Also, from Fig. \ref{c1}, it is observed that, when the environment-induced interatomic interaction is neglected, the  maximum of concurrence generated during evolution may increase with acceleration, in contrast to the expectation that it should decrease monotonically with the Unruh temperature.
This phenomenon is named as the anti-Unruh phenomenon in terms of the entanglement generated in Ref. \cite{zhou21}, in analogy to the anti-Unruh phenomenon in terms of the excitation rate \cite{anti1,anti2}. Remarkably,
the anti-Unruh phenomenon however disappears when the environment-induced interatomic interaction is considered.
In Fig. \ref{c2}, it is shown that the maximal concurrence generated during evolution decreases monotonically as the interatomic separation increases when the environment-induced interatomic interaction is considered. In contrast, it shows an oscillatory behavior in the neglect of the environment-induced interatomic interaction when the separation is much smaller than the transition wavelength.


\subsection{Entanglement degradation}

In the following, we assume that the two-atom system is initially prepared in an entangled state, and investigate the phenomenon of entanglement degradation.
As has been discussed, the only density matrix elements affected by the environment-induced interatomic interaction are $\rho_{AS}$ and $\rho_{SA}$, so we choose the initial state of the two-atom system to be a superposition state of the antisymmetric state $|A\rangle$ and the symmetric state $|S\rangle$, i.e., $\cos\theta|A\rangle+\sin\theta e^{i\varphi}|S\rangle$, where $\theta$ and $\varphi$ are the weight parameter and phase parameter respectively.

\begin{figure}[htbp]
  \centering
  \includegraphics[width=0.32\textwidth]{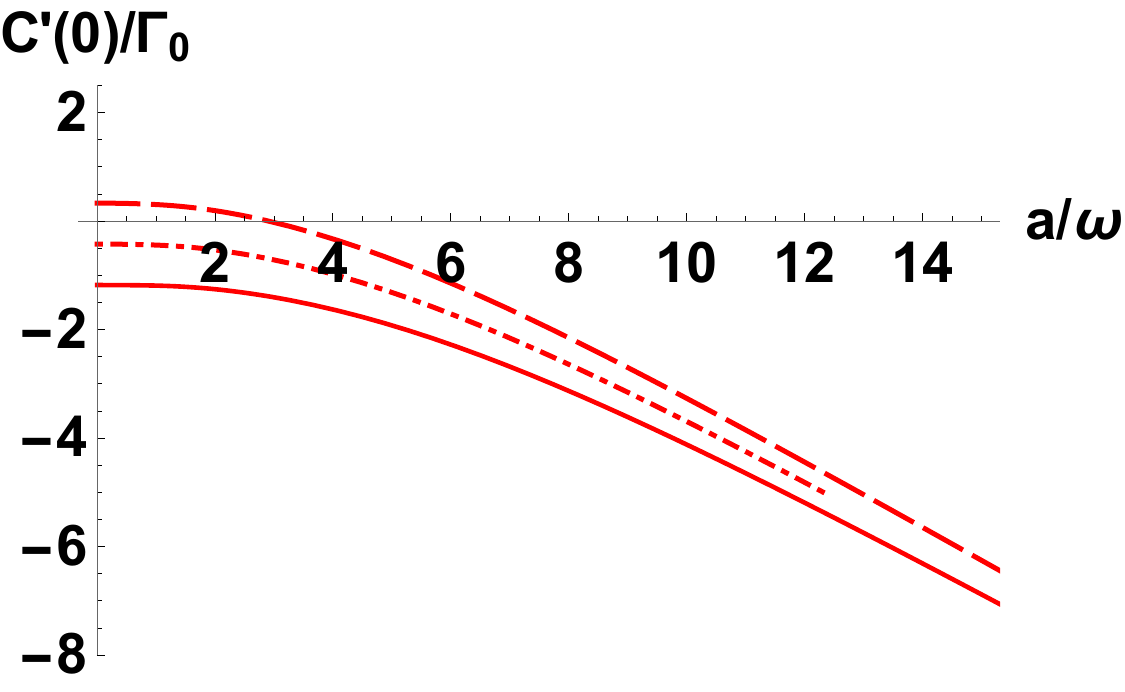}
  \includegraphics[width=0.32\textwidth]{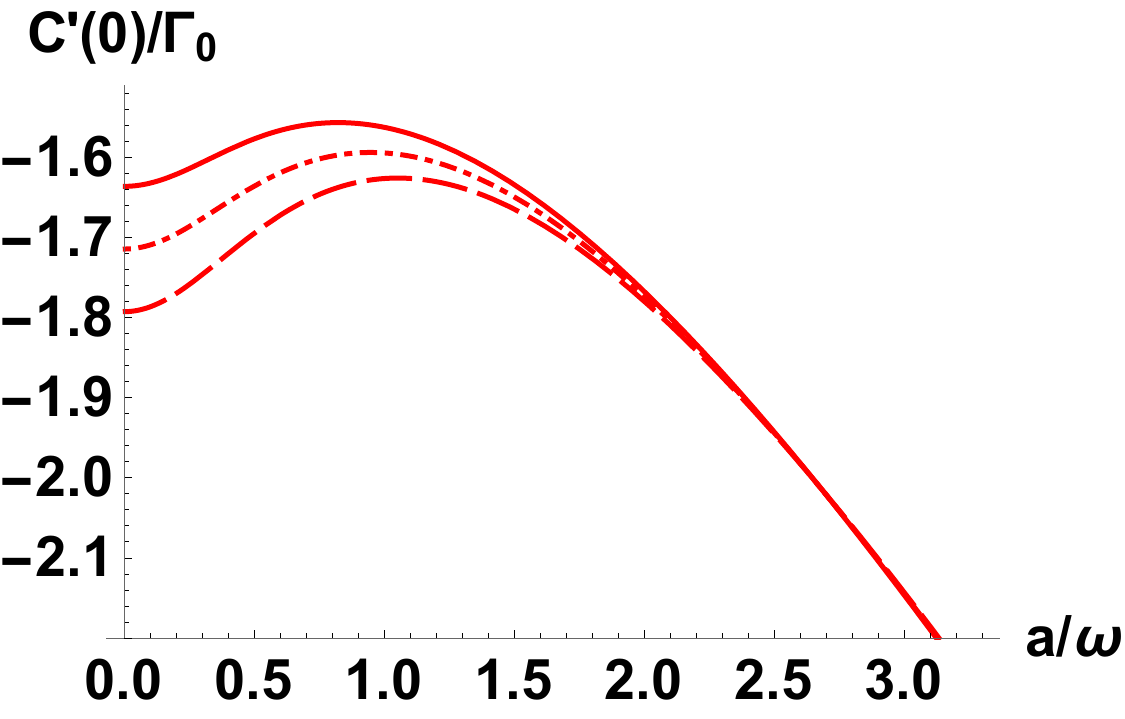}
  \includegraphics[width=0.32\textwidth]{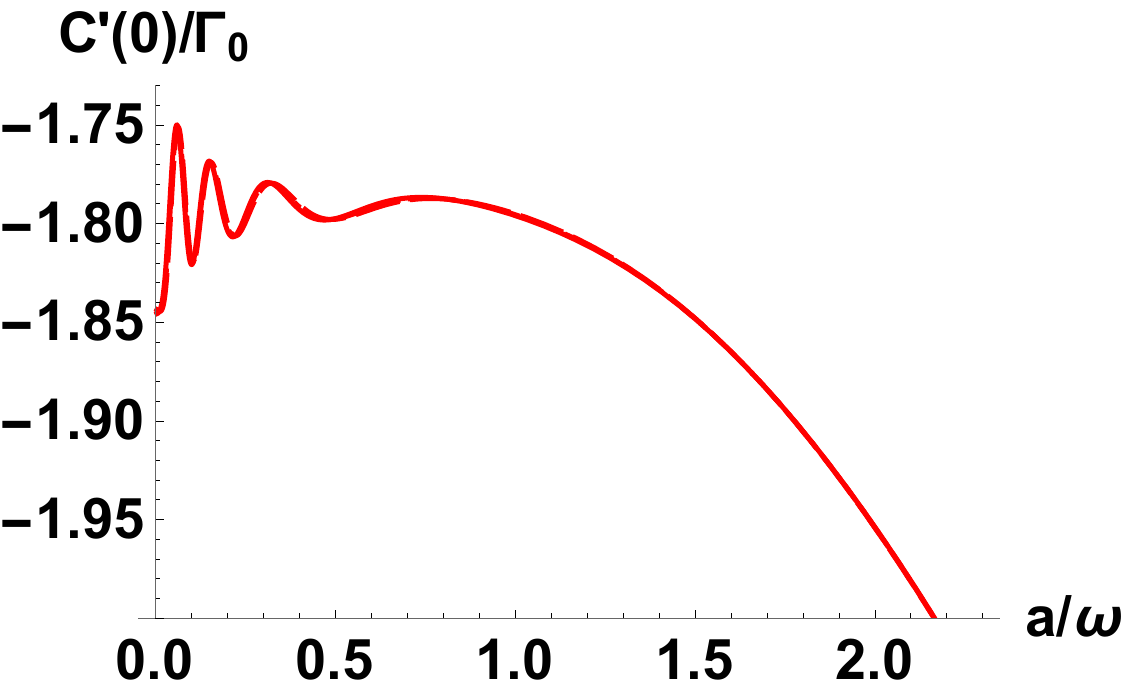}
  \caption{$C'(0)$ as a function of $a/\omega$ with $\omega L=3/10$ (left), $\omega L=3$ (middle), and $\omega L=30$ (right), for uniformly accelerated atoms initially prepared in $\frac{\sqrt{3}}{2}|A\rangle+\frac{1}{2}e^{i\frac{\pi}{4}}|S\rangle$ (solid) and $\frac{\sqrt{3}}{2}|A\rangle+\frac{1}{2}e^{-i\frac{\pi}{4}}|S\rangle$ (dashed). The dot-dashed lines represent those in the case when  the environment-induced  interatomic interaction is neglected.}\label{as}
\end{figure}

\begin{figure}[htbp]
  \centering
  \includegraphics[width=0.32\textwidth]{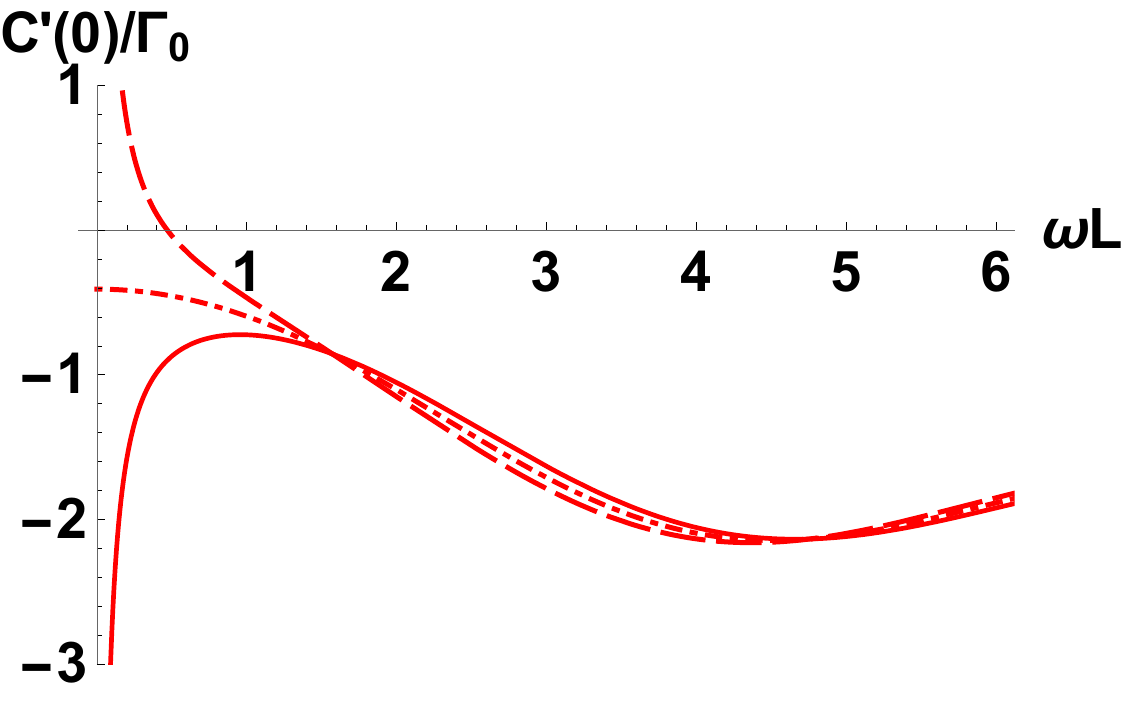}
  \includegraphics[width=0.32\textwidth]{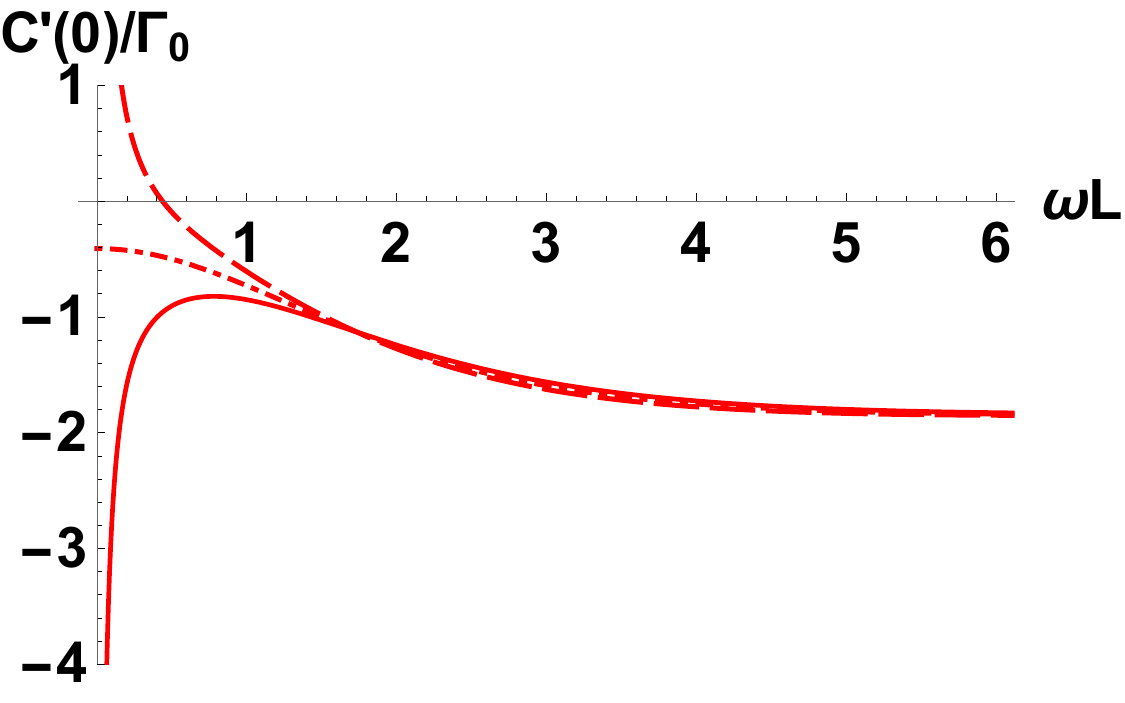}
  \includegraphics[width=0.32\textwidth]{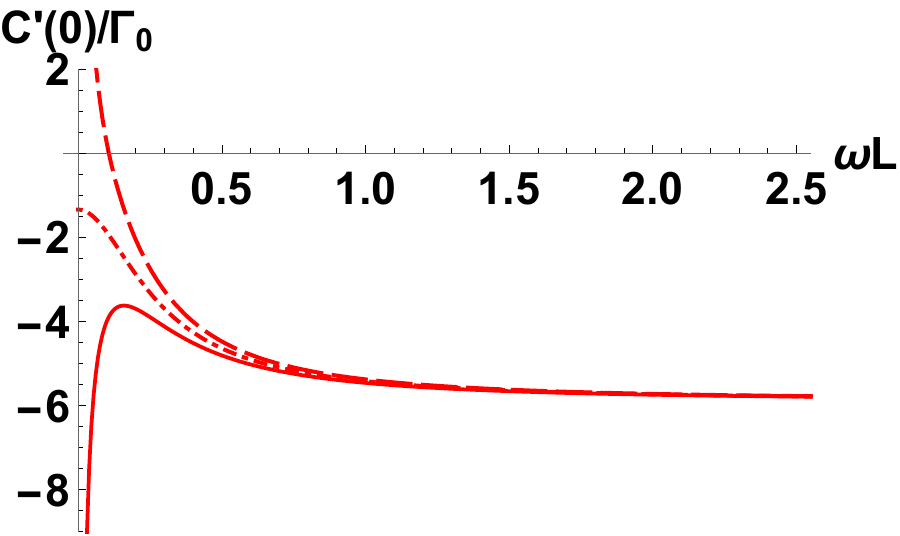}
  \caption{$C'(0)$ as a function of $\omega L$ with $a/\omega=1/10$ (left), $a/\omega=1$ (middle), and $a/{\omega}=10$ (right), for uniformly accelerated atoms initially prepared in $\frac{\sqrt{3}}{2}|A\rangle+\frac{1}{2}e^{i\frac{\pi}{4}}|S\rangle$ (solid) and $\frac{\sqrt{3}}{2}|A\rangle+\frac{1}{2}e^{-i\frac{\pi}{4}}|S\rangle$ (dashed). The dot-dashed lines represent those in the case when the  environment-induced  interatomic interaction is neglected.}\label{as1}
\end{figure}

\begin{figure}[htbp]
  \centering
   \includegraphics[width=0.35\textwidth]{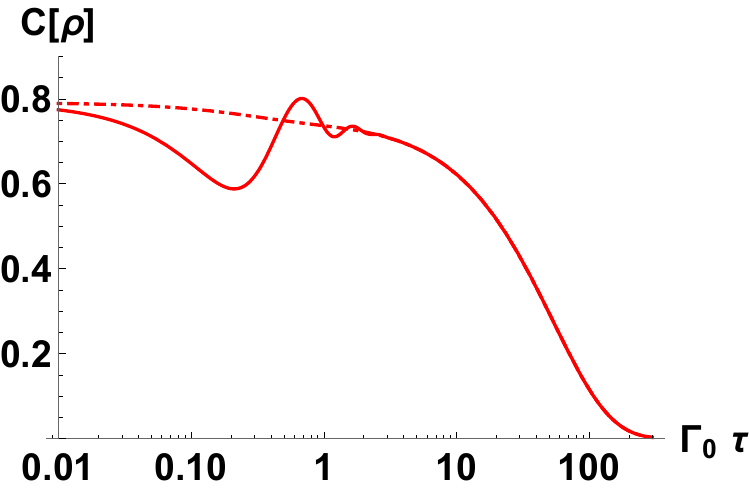}\;\;
  \includegraphics[width=0.35\textwidth]{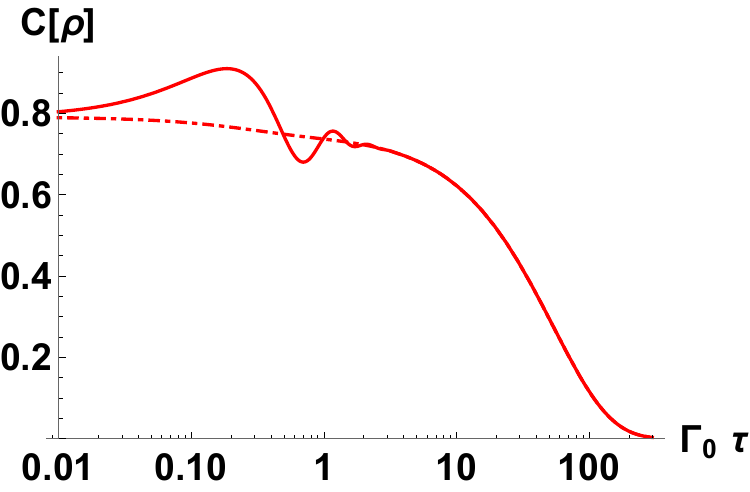}
  \caption{Time evolution of concurrence for uniformly accelerated atoms with (solid) and without (dot-dashed) the environment-induced  interatomic interaction initially prepared in $\frac{\sqrt{3}}{2}|A\rangle+\frac{1}{2}e^{i\frac{\pi}{4}}|S\rangle$ (left), and $\frac{\sqrt{3}}{2}|A\rangle+\frac{1}{2}e^{-i\frac{\pi}{4}}|S\rangle$ (right). Here $a/{\omega}=1/2$, and $\omega L = 3/10 $.}\label{pg2}
\end{figure}

The rate of change of concurrence at the initial time can be calculated as
\bea\label{k2}
\nn {C}'(0)&=&\frac{-4A_{1}[\cos^{2}2\theta+\sin^{2}2\theta\sin^2\varphi]+4A_{2}\cos2\theta
-2D\sin^{2}2\theta\sin2\varphi}
{\sqrt{\cos^{2}2\theta+\sin^{2}2\theta\sin^2\varphi}}\\
&&-4\sqrt{(A_{1}-A_{2}\cos2\theta)^2-(B_{1}-B_{2}\cos2\theta)^2}.
\eea
From Eq. \eqref{k2},  we know that the term related to the phase parameter $\varphi$ does not exist when the environment-induced interatomic interaction is neglected, while it can be  positive or negative when the environment-induced interatomic interaction is  considered.
Therefore, $C'(0)$ may be increased or decreased, and may change its sign when the  environment-induced interatomic interaction is  considered.
As shown in Figs. \ref{as} and \ref{as1}, when the interatomic separation is small compared with the transition wavelength, $C'(0)$ may change from  negative to positive, i.e. the type of entanglement evolution at the initial time  may change from entanglement degradation  to  entanglement enhancement  when the environment-induced interatomic interaction  is considered.
See Fig. \ref{pg2} (right) for an example.
When the interatomic separation is comparable to the transition wavelength of the atoms, the consideration of the environment-induced interatomic interaction affects only quantitatively on the rate of change of concurrence. When the interatomic separation is large compared to the transition wavelength of the atoms, the effects of environment-induced interatomic interaction becomes negligible, as expected.


\section{Conclusion}

In this paper, we have investigated, in the framework of open quantum systems, the effect of the  environment-induced interatomic interaction on the entanglement dynamics of two uniformly accelerated atoms in the Minkowski vacuum.
We have shown that the environment-induced interatomic interaction may have a significant effect on the entanglement dynamics of a two-atom system, depending on the choice of the initial state. The relevant conclusions are summarized as follows.

First, when the two atoms are initially prepared  in a state such that one is in  the ground state and other is in the excited one,  which is separable, it is shown that, compared with  when the environment-induced interaction is neglected,  the parameter space of acceleration and interatomic separation which allows
 entanglement generation is enlarged,  the rate of entanglement generation at the neighborhood of the initial time is enhanced, and the maximum of concurrence generated  during evolution is increased, when the environment-induced interaction between the atoms is taken into account. In this sense,  the environment-induced interatomic interaction assists entanglement generation.

Second, when the atoms are initially prepared in a superposition state of the antisymmetric state  and the symmetric state, which is entangled, the type of entanglement evolution at the initial time  may change from entanglement degradation  to  entanglement enhancement  when the environment-induced interatomic interaction  is considered.  In this sense,  the environment-induced interatomic interaction is also beneficial to entanglement generation.

Finally, it is worth emphasizing that when the environment-induced interatomic interaction is neglected, the rate of entanglement generation at the initial time and the maximal concurrence generated during evolution
exhibit a nonmonotonic behavior as the acceleration varies.
However, they decrease monotonically with the acceleration when the  environment-induced interatomic interaction is taken into account.  As a result,  the anti-Unruh phenomenon in terms of the entanglement generation
disappears when the environment-induced interatomic interaction is considered.

\begin{acknowledgments}
This work was supported in part by the NSFC under Grants No. 11805063, No. 11690034, and No. 12075084, and the Hunan Provincial Natural Science Foundation of China under Grant No. 2020JJ3026.
\end{acknowledgments}

\end{document}